\documentclass{fizik}
\usepackage{hyperref}
\hypersetup{
colorlinks=true,
urlcolor=blue,
citecolor=blue}
\usepackage[all]{xy,xypic}
\usepackage{amsfonts,amssymb,amsmath,amsgen,amsopn,amsbsy,theorem,graphicx,epsfig}
\usepackage{eufrak,amscd,bezier,latexsym,mathrsfs,enumerate,multirow}
\usepackage[utf8]{inputenc}\usepackage[english]{babel}
\usepackage[dvipsnames]{xcolor}
\usepackage[pagewise]{lineno}
\usepackage{mathrsfs}
\usepackage{changes}
\usepackage{float}
\restylefloat{table}

\yil{2021}
\vol{TBA}
\fpage{tba1}
\lpage{tba2}
\doi{10.3906/fiz-2012-6}

\title{
Quasinormal modes of charged fermions in linear dilaton black hole spacetime: Exact frequencies }

\author[SAKALLI and TOKG\"{O}Z HYUSEIN]{
\textbf{\.{I}zzet SAKALLI$^{1}$\thanks{izzet.sakalli@emu.edu.tr}~\href{https://orcid.org/0000-0001-7827-9476}{}, 
G\"{u}ln\.{i}hal TOKG\"{O}Z HYUSEIN$^{1,2}$\href{https://orcid.org/0000-0001-7270-4355}{}}\\
\\ 
$^{1}$Physics Department, Eastern Mediterranean
University, Famagusta, North Cyprus \\ 99628 via Mersin 10, Turkey\\
$^{2}$G\"{u}zeloba Mahallesi, \"{O}rnekk\"{o}y Caddesi, \"{O}rnekk\"{o}y Sitesi, 2993 Ada, 11KA Blok, No: 2, Muratpaşa\\ 07230 Antalya, Turkey\\
\\
\textcolor{red}{Corresponding Author: \.{I}zzet SAKALLI$^{1}$}
\\ [1.8em]

\rec{.201}
\acc{.201}
\finv{..201}
}

\newcommand{\bc}{\begin{center}}
\newcommand{\ec}{\end{center}}

\numberwithin{equation}{section}

\renewcommand{\phi}{\varphi}

\begin{document}

\maketitle

\begin{abstract}We study charged massless fermionic perturbations in the background of $4$-dimensional linear dilaton black holes in Einstein-Maxwell-dilaton 
theory with double Liouville-type potentials. We present the analytical fermionic quasinormal modes, whose Dirac equations are solved in terms of hypergeometric functions. We also discuss the stability of these black holes under the charged fermionic perturbations.

\end{abstract}

\section{Introduction}

At high enough energies, it is quite possible that gravity is not described by the action of Einstein's general relativity theory. Today, the literature has a compelling evidence that in string theory (ST), gravity becomes a scalar-tensor \cite{isp1}. The low-energy limit of the ST corresponds to Einstein's
gravitational attraction, which is non-minimally coupled to a scalar dilaton
field \cite{isp2}. When the Einstein-Maxwell theory is combined with a dilaton
field, the resulting spacetime solutions have important consequences. There
have been significant studies in the literature to find exact solutions of the
Einstein-Maxwell-dilaton (EMD) theory. For example, in the absence of a dilaton potential, EMD
gravity's charged dilaton black hole (BH) solutions were found by many
researchers \cite{isp3,isp4,isp6,isp7,isp8,isp9,isp10,isp11,isp12}. Essentially, the presence of the dilaton alters the
arbitrary structure of spacetime and causes to curvature singularities with
finite radii. The obtained dilaton BHs in general have non-asymptotically flat (NAF) structure. In recent years, even in the various extensions of general relativity, there has been an active interest in the NAF spacetimes. Among them, probably the
most important ones are the asymptotic anti-de Sitter (AdS) BHs \cite{isp13}, which
play a vital role in string and quantum gravity theories as well as for the
AdS/CFT (conformal field theory) correspondence, which is sometimes called
Maldacena duality or gauge/gravity duality \cite{isp14}.

The BH solutions of EMD theory were studied by many researchers: the uncharged
EMD BH solutions were found in \cite{isp15,isp16,isp17}, while the charged EMD BH solutions were
considered in \cite{isp18,isp19}. With the inclusion of the Liouville-type potentials, the static charged BH solutions were found by \cite{isp20,isp21,isp22}. The
generalization to dyonic (having both electric and magnetic charges) BH
solutions in $4$-dimensional and higher-dimensional EMD gravity with single
and double Liouville-type potentials were also found in \cite{isp23}. In fact, similar
to the Higgs potential, the double Liouville-type potentials have the ability
to admit local extremes and critical points. However, a single Liouville
potential lacks from these features. Besides, the double Liouville-type
potentials also appear when the higher-dimensional theories are compressed
into a $4$-dimensional spacetime. Many of the BH solutions with the
double Liouville-type potentials to date can be seen from \cite{isp24} and references therein. Among them
Mazharimousavi et al's BH \cite{isp25} in the EMD theory with the double
Liouville-type potentials has some unique features. First of all, those BHs
cover Reissner-Nordstr\"{o}m (RN) type BHs and
Bertotti-Robinson (BR) spacetimes \cite{isp26} interpolated within the same metric.
Remarkably, in between the two spacetimes, there exists a linear dilaton BH
(LDBH) \cite{isp26n} solution for the specific values of the dilaton, double Liouville-type
potentials, and EMD parameters. The particular motivation of this work is to
compute the quasinormal modes (QNMs) \cite{isp27,isp28} for the charged fermionic field perturbations \cite{isp29,isp30,isp31,isp32} in the
spacetime of those LDBHs. Since the QNMs can signal information about the
stability of BHs, we shall also analyze the stability of the LDBHs under the
charged fermionic perturbations.

Meanwhile, we would like to remind the reader for information purposes that
QNMs are nothing but the energy dissipation of a perturbed BH. Similar to an
ordinary object, when a BH is perturbed it begins to ring with its natural
frequencies, which are the modes of its energy dissipation. These
characteristic modes are called QNMs. The amplitudes of their oscillations
decay in time. The amplitude of the oscillation can be approximated by
\begin{align}
\Psi & \approx e^{i\omega t}\text{ }=e^{-\omega_{I}t}\cos(\omega
_{R}t),\label{is0}%
\end{align}
where $\omega=\omega_{R}+i\omega_{I}$ is the frequency of the QNM \cite{isp28}. Here,
$\omega_{R} $ and $\omega_{I}$\ are the frequencies of oscillatory and
exponential modes, respectively. After LIGO's great successes about the
gravitational wave measurements \cite{isp33}, the subject of QNMs has gained considerable
importance in the direct identification of BHs. Today, there are numerous
well-known studies which show that the surrounding geometry of a BH
experiences QNMs under perturbations (see for example
\cite{isp34,isp35,isp36,isp37,isp38,isp39,isp40,isp40n} and references therein).
One of the most important features to know about QNMs is that they allow us to analyze the quantum entropy/area spectrum of the BHs. In this regard, the reader is
referred to \cite{isp41,isp42,isp43,isp44,isp45,isp46,isp47} and references therein.

The organization of the present paper is as follows: In Sec. \ref{sec2}, we
review the LDBH solution, which is obtained from the $4$-dimensional action of
the EMD theory having the double Liouville-type potentials. In Sec.
\ref{sec3}, we first derive the spin-$\frac{1}{2}$ field equations by using
$4$-dimensional charged massless Dirac equations within the framework of
Newman-Penrose (NP) formalism. Solution procedure to the obtained Dirac
equations is given in Sec. \ref{sec4}. Section \ref{sec5} is devoted to the
computations of the QNM frequencies of the charged fermions. We summarize our
results in Sec. \ref{sec6}. (Throughout the paper, we use the geometrized
units of $G=c=\hbar=1$.)

\section{LDBH spacetime in EMD theory with double Liouville-type potential}

\label{sec2} $4$-dimensional action of the EMD theory having the double
Liouville-type potentials is given by \cite{isp25}
\begin{equation}
S=\int\mathrm{d}^{4} x \sqrt{-g}\left( \frac{1}{2} R-\frac{1}{2} \partial
_{\mu} \phi\partial^{\mu} \phi-V(\phi)-\frac{1}{2} W(\phi)\left(
F_{\lambda\sigma} F^{\lambda\sigma}\right) \right) ,\label{is1}%
\end{equation}
where
\begin{equation}
V(\phi)=V_{1} \mathrm{e}^{\beta_{1} \phi}+V_{2} \mathrm{e}^{\beta_{2} \phi},
\text{ \ \ \ \ } W(\phi)=\lambda_{1} \mathrm{e}^{-2 \gamma_{1} \phi}%
+\lambda_{2} \mathrm{e}^{-2 \gamma_{2} \phi},
\end{equation}
\label{is2}
where $\phi$ is the dilaton, $V_{1,2}$ are the double
Liouville-type potentials, $\gamma_{1,2}$ denote the dilaton parameter, and
$\lambda_{1,2},\thinspace\beta_{1,2}$ are constants. Moreover, $R$ is the
Ricci scalar and the Maxwell 2-form is given by%
\begin{equation}
F=d\mathcal{A}.\label{5}%
\end{equation}
For the choice of pure magnetic potential
\begin{equation}
\mathcal{A}=-Q\cos\theta d\varphi,\label{6}%
\end{equation}
where $Q$ denotes the charge, we have\newline%
\begin{equation}
F=Q\sin\theta d\theta\wedge d\varphi.\label{7}%
\end{equation}
It is worth noting that with the current choice of $W(\phi)$, the
magnetic-electric symmetry, which exists in the standard dilatonic coupling,
namely $\lambda_{1}\left( \lambda_{2}\right) =0,$ is not valid anymore. Thus,
as highlighted in \cite{isp25}, the charged LDBH spacetime is pure magnetic.
Variations of the action (\ref{is1}) with respect to the gravitational field
$g_{\mu\nu}$ and the dilaton $\phi$ yield the following EMD field equations
\begin{equation}
R_{\mu\nu}=\partial_{\mu}\phi\partial_{\nu}\phi+Vg_{\mu\nu}+W\left(
2F_{\mu\lambda}F_{\nu}^{\lambda}-\frac{1}{2}F_{\lambda\sigma}F^{\lambda\sigma
}g_{\mu\nu}\right)  ,\label{8}%
\end{equation} 
\begin{equation}
\nabla^{2}\phi-V^{\prime}-\frac{1}{2}W^{\prime}\left(  F_{\lambda\sigma
}F^{\lambda\sigma}\right)  =0,\label{9}%
\end{equation}
where $R_{\mu\nu}$ is the Ricci tensor and the prime symbol ($^{\prime}$)
denotes the derivative with respect to $\phi$. Furthermore, the Maxwell
equation is obtained with the variation with respect to $\mathcal{A}$ as
\begin{equation}
d\left(  W^{\star}F\right)  =0\label{10}%
\end{equation}
in which the Hodge star ($^{\star}$) refers to duality. After substituting the
following metric (ansatz):
\begin{equation}
ds^{2}=B(r)dt^{2}-\frac{dr^{2}}{B(r)}-R(r)d\Omega_{2}^{2},\label{11}%
\end{equation}
into Eqs.(\ref{8}) and (\ref{9}) and in the sequel of compelling calculations as made in \cite{isp25}, the field equations result in
the metric functions of the LDBH as follows%
\begin{equation}
B(r)=b\frac{(r-r_{2})(r-r_{1})}{r}\text{ \ \ and \ \ }R(r)=A^{2}%
r\text{,}\label{12}%
\end{equation}
where
\begin{equation}
b=\frac{1}{A^{2}}-2(V_{1}+V_{2})>0,\text{ \ \ \ and \ \ }A^{2}=2\lambda
_{2}Q^{2},\text{ \ \ \ }(A:Real\medspace constant).\label{13}%
\end{equation}
The inner and outer horizons of the LDBH are given by \cite{isp25}
\begin{equation}
r_{1}=\frac{1}{2b}\left(  c-\sqrt{c^{2}-4ab}\right)  \text{ \ and \ \ \ }%
r_{2}=\frac{1}{2b}\left(  c+\sqrt{c^{2}-4ab}\right)  ,\label{14}%
\end{equation}
in which the physical parameters read
\begin{equation}
c=4M\text{ \ \ \ \ and \ \ }a=\frac{\lambda_{1}}{\lambda_{2}A^{2}},\label{15}%
\end{equation}
where $M$ denotes the quasilocal mass \cite{isp48} and 
\begin{equation}
\phi\left(  r\right)  =-\frac{1}{\sqrt{2}}\ln\left(  r\right)  ,\text{ }%
\beta_{1}=\beta_{2}=\sqrt{2},\text{\ \ }\gamma_{1}=-\frac{1}{\sqrt{2}}\text{
},\text{ }\gamma_{2}=\frac{1}{\sqrt{2}},\text{\ \ \ }\label{16}%
\end{equation}

\begin{equation}
V=\frac{V_{1}+V_{2}}{r},\text{ \ \ and \ \ \ }W=\frac{\lambda_{1}}{r}%
+\lambda_{2}r.\label{17}%
\end{equation}
The above spacetime corresponds to a phase transition geometry, which changes
the structure of spacetime from RN to BR \cite{isp25}. The case of $c^{2}=4ab$ gives us the
extremal ($r_{2}=r_{1}$) LDBHs whose congenerics can be seen in
\cite{isp26n,isp49}.

\section{\bigskip Charged Dirac equation in LDBH geometry}

\label{sec3} In the NP formalism \cite{isp50}, massless Dirac equations with
charge coupling are given as follows \cite{isp29,isp51}%
\begin{align}
\left[D+iql^{j}\mathcal{A}_{j}+\varepsilon-\rho\right]  F_{1}+\left[
\overline{\delta}+iq\overline{m}^{j}\mathcal{A}_{j}+\pi-\alpha\right]  F_{2}
& =0,\nonumber\\
\left[  \delta+iqm^{j}\mathcal{A}_{j}^{\text{ \ \ }}+\beta-\tau\right]
F_{1}+\left[  \Delta+iqn^{j}\mathcal{A}_{j}^{\text{ \ \ }}+\mu-\gamma\right]
F_{2}  & =0,\nonumber\\
\left[  D+iql^{j}\mathcal{A}_{j}+\overline{\varepsilon}-\overline{\rho
}\right]  \widetilde{G}_{2}-\left[  \delta+iqm^{j}\mathcal{A}_{j}%
+\overline{\pi}-\overline{\alpha}\right]  \widetilde{G}_{1}  & =0,\nonumber\\
\left[  \Delta+iqn^{j}\mathcal{A}_{j}^{\text{ \ \ }}+\overline{\mu}%
-\overline{\gamma}\right]  \widetilde{G}_{1}-\left[  \overline{\delta
}+iq\overline{m}^{j}\mathcal{A}_{j}+\overline{\beta}-\overline{\tau}\right]
\widetilde{G}_{2}  & =0,\label{18}%
\end{align}
where $q$ is the charge of the fermion and $A_{j}$ represents the $j^{th}$
component of the vector potential of the background electromagnetic field
(\ref{6}). The wave functions $F_{1},F_{2},\widetilde{G}_{1},\widetilde{G}%
_{2}$ \ represent the Dirac spinors while $\alpha,\beta,\gamma,\varepsilon
,\mu,\pi,\rho,\tau$ are the spin (Ricci rotation) coefficients. The
directional derivatives for NP tetrads are defined as
\begin{equation}
D= l^{j}\nabla_{j}, \medspace \Delta=n^{j}\nabla_{j},
\medspace \delta= m^{j}\nabla_{j}, \medspace \bar{\delta}=\bar
{m}^{j}\nabla_{j},\label{19}%
\end{equation}
In the meantime, a bar over a quantity stands for the complex conjugation. We
choose a complex null tetrad $\left\{  l,n,m,\overline{m}\right\}  $ (the
covariant one-forms) for the LDBH geometry as
\begin{align}
l_{j}  & =\frac{1}{\sqrt{2}}\left[  \sqrt{B(r)},-\frac{1}{\sqrt{B(r)}%
},0,0\right]  ,\nonumber\\
n_{j}  & =\frac{1}{\sqrt{2}}\left[  \sqrt{B(r)},\frac{1}{\sqrt{B(r)}%
},0,0\right]  ,\nonumber\\
m_{j}  & =A\sqrt{\frac{r}{2}}\left[  0,0,1,i\sin\theta\right]  ,\nonumber\\
\overline{m}_{j}  & =A\sqrt{\frac{r}{2}}\left[  0,0,1,-i\sin\theta\right]
,\label{20}%
\end{align}
which all together satisfy the orthogonality conditions: $l.n=-m.\overline
{m}=1$. The non-zero spin coefficients are obtained as
\begin{align}
\rho & =\mu=-\frac{-1}{2\sqrt{2}}\frac{\sqrt{B(r)}}{r},\nonumber\\
\epsilon & =\gamma=\frac{b}{4\sqrt{2B(r)}}\left(  1-\frac{r_{2}r_{1}}{r^{2}%
}\right)  ,\nonumber\\
\alpha & =-\beta=\frac{\cot\theta}{2A\sqrt{2r}}.\label{21}%
\end{align}
The form of the Dirac equations (\ref{18}) suggests that the spinors can be
chosen as follows
\begin{align}
F_{1}  & =f_{1}\left(  r\right)  A_{1}\left(  \theta\right)  e^{i\left(
kt+m\varphi\right)  },\nonumber\\
\widetilde{G}_{1}  & =g_{1}\left(  r\right)  A_{2}\left(  \theta\right)
e^{i\left(  kt+m\varphi\right)  },\nonumber\\
F_{2}  & =f_{2}\left(  r\right)  A_{3}\left(  \theta\right)  e^{i\left(
kt+m\varphi\right)  },\nonumber\\
\widetilde{G}_{2}  & =g_{2}\left(  r\right)  A_{4}\left(  \theta\right)
e^{i\left(  kt+m\varphi\right)  },\label{22}%
\end{align}
where $k$ is the frequency of the wave corresponding to the Dirac particle and
$m$ is the azimuthal quantum number.

\bigskip

\section{\bigskip Solution of charged Dirac equation in LDBH geometry}

\label{sec4}

After substituting the\ spin coefficients (\ref{21}) and the spinors
(\ref{22}) into the Dirac equations (\ref{18}), one gets
\begin{align}
\frac{1}{f_{2}}\widetilde{Z}f_{1}-\frac{\left(  LA_{3}\right)  }{A_{1}}  &
=0,\nonumber\\
\frac{1}{f_{1}}\overline{\widetilde{Z}}f_{2}+\frac{\left(  L^{\dag}%
A_{1}\right)  }{A_{3}}  & =0,\nonumber\\
\frac{1}{g_{1}}\widetilde{Z}g_{2}-\frac{\left(  L^{\dag}A_{2}\right)  }%
{A_{4}}  & =0,\nonumber\\
\frac{1}{g_{2}}\overline{\widetilde{Z}}g_{1}+\frac{\left(  LA_{4}\right)
}{A_{2}}  & =0.\label{23}%
\end{align}
The radial operators appear in the above simplified Dirac equations are
\begin{align}
\widetilde{Z}  & =A\sqrt{\Lambda}\partial_{r}+H+\frac{iAkr}{\sqrt{\Lambda}%
},\nonumber\\
\overline{\widetilde{Z}}  & =A\sqrt{\Lambda}\partial_{r}+H-\frac{iAkr}%
{\sqrt{\Lambda}},\label{24}%
\end{align}
in which
\begin{equation}
\Lambda=b(r-r_{2})(r-r_{1}),\label{25}%
\end{equation}

\begin{equation}
H=\frac{A}{2r}\left[  \frac{b(r^{2}-r_{2}r_{1})}{2\sqrt{\Lambda}}%
+\sqrt{\Lambda}\right]  .\label{26}%
\end{equation}
The angular operators are
\begin{align}
L  & =\partial_{\theta}+\frac{m}{\sin\theta}+\left(  \frac{1}{2}-p\right)
\cot\theta,\nonumber\\
L^{\dag}  & =\partial_{\theta}-\frac{m}{\sin\theta}+\left(  \frac{1}%
{2}+p\right)  \cot\theta,\label{27}%
\end{align}
where $p=qQ$. Further, choosing
\begin{equation}
f_{1}=g_{2},\medspace f_{2}=g_{1},\medspace A_{1}=A_{2},\medspace
A_{3}=A_{4},\label{28}%
\end{equation}
and introducing a real eigenvalue $\lambda$ which is the separation constant
of the complete Dirac equations, one can separate the Dirac equations into two
sets of master equations. The radial master equations read
\begin{align}
\overline{\widetilde{Z}}g_{1}  & =-\lambda g_{2},\nonumber\\
\widetilde{Z}g_{2}  & =-\lambda g_{1},\label{29}%
\end{align}
and the angular master equations become
\begin{align}
L^{\dag}A_{2}  & =\lambda A_{4},\nonumber\\
LA_{4}  & =-\lambda A_{2}.\label{30}%
\end{align}

\subsection{\bigskip Angular equation}

The laddering operators $L$ and $L^{\dag}$ \cite{lee} govern the spin-weighted
spheroidal harmonics as
\begin{align}
\left(  \partial_{\theta}-\frac{m}{\sin\theta}-s\cot\theta\right)  \left(
_{s}Y_{l}^{m}(\theta)\right)   & =-\sqrt{\left(  l-s\right)  \left(
l+s+1\right)  }_{s+1}Y_{l}^{m}\left(  \theta\right)  ,\nonumber\\
\left(  \partial_{\theta}+\frac{m}{\sin\theta}+s\cot\theta\right)  \left(
_{s}Y_{l}^{m}(\theta)\right)   & =\sqrt{\left(  l+s\right)  \left(
l-s+1\right)  }_{s-1}Y_{l}^{m}\left(  \theta\right)  .\label{31}%
\end{align}
Eigenfunctions $_{s}Y_{l}^{m}(\theta)$, called the spin-weighted spheroidal
harmonics, are complete and orthogonal, and have an explicit form
\cite{goldberg}:
\begin{align}
_{s}Y_{l}^{m}(\theta)  & =\sqrt{\frac{2l+1}{4\pi}\frac{\left(  l+m\right)
!\left(  l-m\right)  !}{\left(  l+s\right)  !\left(  l-s\right)  !}}\left(
\sin\frac{\theta}{2}\right)  ^{2l}\nonumber\\
& \times\overset{l}{\underset{r=-l}{\sum\left(  -1\right)  }}^{l+m-r}%
\binom{l-s}{r-s}\binom{l+s}{r-m}\left(  \cot\frac{\theta}{2}\right)
^{2r-m-s},\label{32}%
\end{align}
where $l$ and $s$ are the angular quantum number and the spin-weight,
respectively. They satisfy the following expressions:
\begin{equation}
l=\left\vert s\right\vert ,\left\vert s\right\vert +1,\left\vert s\right\vert
+2,.... \medspace \text{and }-l<m<+l.\label{33}%
\end{equation}
Comparison between the angular master equations (\ref{30}) and (\ref{31})
leads us to identify
\begin{align}
A_{2}  & =_{-(\frac{1}{2}+p)}Y_{l}^{m},\nonumber\\
A_{4}  & =_{(\frac{1}{2}-p)}Y_{l}^{m}.\label{34}%
\end{align}
Since $l$ and $s$ both must be integers or half-integers, we impose the "Dirac
quantization condition" \cite{dirac}:%

\begin{equation}
2qQ=2p=n,\text{\ \ \ \ \ \ }n=0,\pm1,\pm2....,\label{35}%
\end{equation}
and obtain the eigenvalue of the spin-weighted spheroidal harmonic equation
as
\begin{align}
\lambda & =-\sqrt{\left(  l+\frac{1}{2}\right)  ^{2}-p^{2}}%
,\text{\ \ \ \bigg(Real: $\left(  l+\frac{1}{2}\right) \geq p$\bigg),}%
\nonumber\\
\therefore\medspace \medspace \lambda^{2}  & =\left(  l+\frac{1}{2}\right)
^{2}-p^{2}.\label{36}%
\end{align}

\subsection{\bigskip Radial equation and Zerilli potential}

\begin{figure}[h]
\centering
\includegraphics[width=12cm,height=8cm]{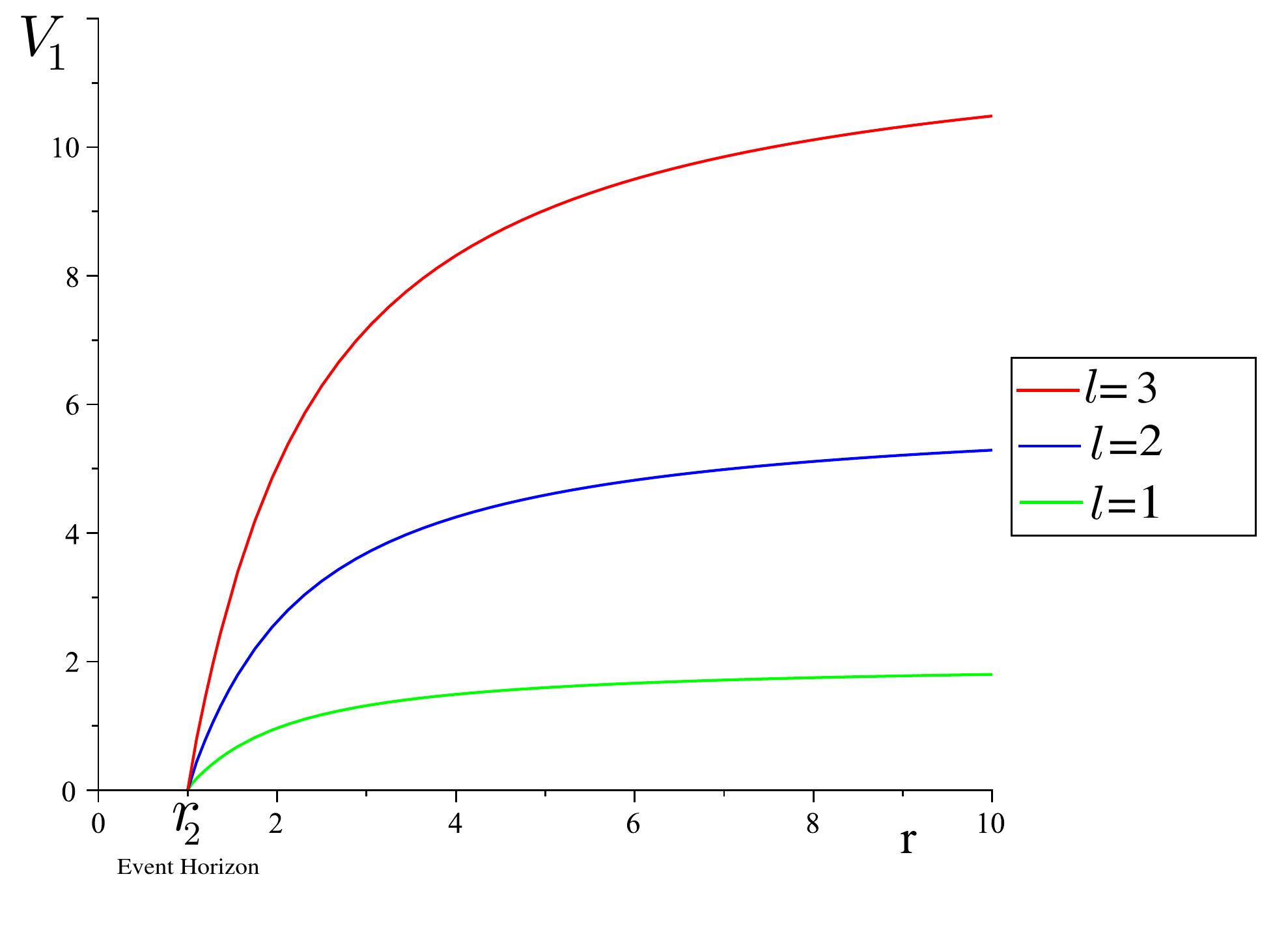}
\caption{$V_{1}$ versus $r$ graph. The physical parameters are chosen as
$A=b=r_{2}=2r_{1}=2p=1$. Plot is governed by Eq. (\ref{43}).} \label{fig1}
\end{figure}

\begin{figure}[h]
\centering
\includegraphics[width=12cm,height=8cm]{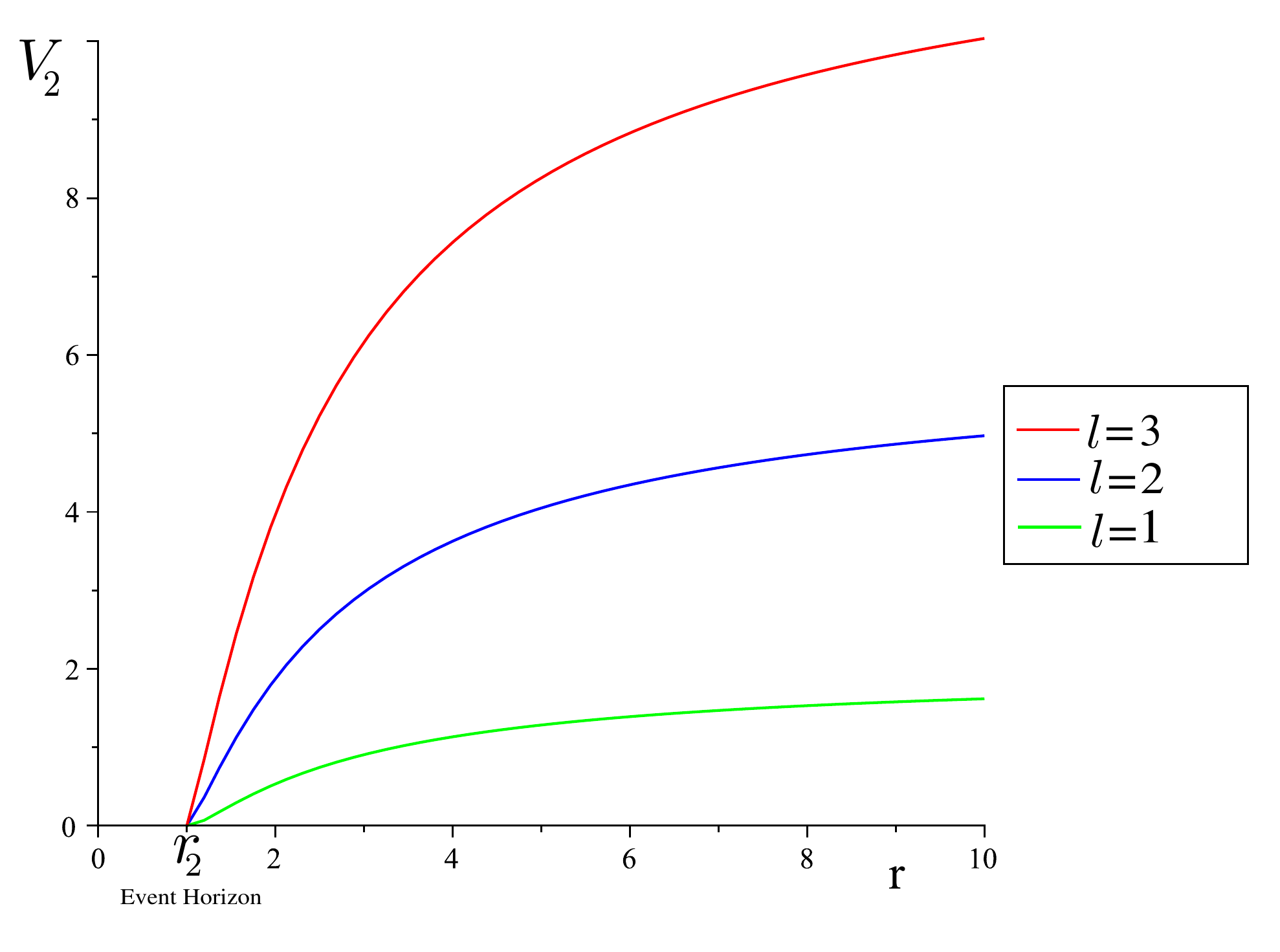}
\caption{$V_{2}$ versus $r$ graph. The physical parameters are chosen as
$A=b=r_{2}=2r_{1}=2p=1$. Plot is governed by Eq. (\ref{43}).} \label{fig2}
\end{figure}

Letting
\begin{align}
g_{1}  & =\frac{G_{1}}{A(r\Lambda)^{\frac{1}{4}}},\nonumber\\
\ g_{2}  & =\frac{G_{2}}{A(r\Lambda)^{\frac{1}{4}}},\label{37}%
\end{align}
and substituting them into the radial master equations (\ref{29}), we obtain
\begin{align}
G_{1}^{\prime}-\frac{ik}{B}  & =-\frac{\lambda G_{2}}{A\sqrt{\Lambda}%
},\nonumber\\
G_{2}^{\prime}+\frac{ik}{B}  & =-\frac{\lambda G_{1}}{A\sqrt{\Lambda}%
}.\label{38}%
\end{align}
Introducing the tortoise coordinate \cite{isp26n}
\begin{equation}
dr_{\ast}=\frac{dr}{B}\text{ \ }\longrightarrow\text{\ \ \ }r_{\ast}=\frac
{1}{b(r_{2}-r_{1})}\ln\left[  \frac{\left(  r-r_{2}\right)  ^{r_{2}}}{\left(
r-r_{1}\right)  ^{r_{1}}}\right]  ,\label{39}%
\end{equation}
one can rewrite Eq. (\ref{38}) as%
\begin{align}
G_{1,r_{\ast}}-ik  & =-\frac{\lambda\sqrt{\Lambda}G_{2}}{Ar},\nonumber\\
G_{2,r\ast}+ik  & =-\frac{\lambda\sqrt{\Lambda}G_{1}}{Ar}.\label{40}%
\end{align}
Setting the solutions of the above equations into the following forms
\begin{align}
G_{1}  & =P_{1}+P_{2},\nonumber\\
G_{2}  & =P_{1}-P_{2},\label{41}%
\end{align}
we get two decoupled radial equations, which correspond to $1$-dimensional Schr\"{o}dinger equation or the so-called Zerilli equation \cite{isp26n} :
\begin{equation}
P_{j,r\ast r\ast}+k^{2}P_{j}=V_{j}P_{j},\text{ \ \ \ \ \ }j=1,2\label{42}%
\end{equation}
where the effective potentials are given by
\begin{equation}
V_{j}=\lambda^{2}\frac{B}{R}+(-1)^{j}\frac{\lambda b}{2r^{2}}\sqrt{\frac{B}%
{R}}\left[  r(r_{1}+r_{2})-2r_{2}r_{1}\right]  .\label{43}%
\end{equation}
The near-horizon and asymptotic limits of the potentials are as
follows
\begin{align}
\bigskip\lim_{r\rightarrow r_{2}}V_{j}  & \equiv\lim_{r^{\ast}\rightarrow
-\infty}V_{j}=0,\nonumber\\
\lim_{r\rightarrow\infty}V_{j}  & \equiv\lim_{r^{\ast}\rightarrow\infty}%
V_{j}=\frac{\lambda^{2}b}{A^{2}}.\label{44}%
\end{align}
For the outer region of LDBH, the behaviors of potentials $V_{j=1,2}$ according to the various angular quantum numbers are depicted in Figures \ref{fig1} and \ref{fig2}, respectively.

Near-horizon ($r\rightarrow r_{2}$) solutions of the radial equations
(\ref{42}) read
\begin{equation}
P_{j}=C_{1j}e^{ikr_{\ast}}+C_{2j}e^{-ikr\ast}, \label{45}%
\end{equation}
and thus we have%
\begin{align}
G_{1}  & =P_{1}+P_{2}\text{ \ \ }\longrightarrow\text{\ \ \ }G_{1}%
=D_{1}e^{-ikr\ast}+D_{3}e^{ikr_{\ast}},\label{46}\\
G_{2}  & =P_{1}-P_{2}\text{ \ \ }\longrightarrow\text{\ \ \ }G_{2}%
=D_{2}e^{-ikr\ast}+D_{4}e^{ikr_{\ast}}.\nonumber
\end{align}
We impose one of the QNM\ conditions that the wave should be purely ingoing at
the horizon by letting $D_{3}=D_{4}=0$; thence%

\begin{equation}
G_{j}=D_{j}e^{-ikr\ast}.\label{47}%
\end{equation}
Asymptotic ($r\rightarrow\infty$) solutions of the radial equations (\ref{42})
are found as%
\begin{equation}
P_{j}=\widetilde{C}_{1j}e^{i\eta r_{\ast}}+\widetilde{C}_{2j}e^{-i\eta r\ast
},\label{48}%
\end{equation}
where%
\begin{equation}
\eta=\frac{1}{A}\sqrt{k^{2}A^{2}-\lambda^{2}b}.\label{49}%
\end{equation}
For $k^{2}>\frac{\lambda^{2}b}{A^{2}}$, the Dirac waves can propagate to
spatial infinity without fading. The asymptotic solutions to the radial
equations are then found to be
\begin{align}
G_{1}  & =P_{1}+P_{2}\text{ \ \ }\longrightarrow\text{\ \ \ }G_{1}%
=\widetilde{D}_{1}e^{i\eta r_{\ast}}+\widetilde{D}_{3}e^{-i\eta r\ast
},\nonumber\\
G_{2}  & =P_{1}-P_{2}\text{ \ \ }\longrightarrow\text{\ \ \ }G_{2}%
=\widetilde{D}_{2}e^{i\eta r_{\ast}}+\widetilde{D}_{4}e^{-i\eta r\ast
}.\label{50}%
\end{align}
We impose the other QNM condition which requires the propagating waves to be purely outgoing at spatial
infinity. To this end, we simply set $\widetilde{D}_{3}=\widetilde{D}_{4}=0$
in Eq. (\ref{50}) and obtain%
\begin{equation}
G_{j}=\widetilde{D}_{j}e^{i\eta r_{\ast}}.\label{51}%
\end{equation}
At the asymptotic region, the tortoise coordinate becomes
\begin{equation}
r_{\ast}=\frac{1}{b(r_{2}-r_{1})}\ln\left[  \frac{\left(  r-r_{2}\right)
^{r_{2}}}{\left(  r-r_{1}\right)  ^{r_{1}}}\right]  \approx\frac{1}%
{b(r_{2}-r_{1})}\ln r^{r_{2}-r_{1}}.\label{52}%
\end{equation}
Therefore, one can see that
\begin{equation}
\eta r_{\ast}\approx\frac{\widetilde{\alpha}}{r_{2}-r_{1}}\ln r^{r_{2}-r_{1}%
}\text{, \ \ \ \ }\label{53}%
\end{equation}
where $\widetilde{\alpha}=\frac{\eta}{b}$ and $e^{i\eta r_{\ast}%
}=r^{i\widetilde{\alpha}}$. By this way, the asymptotic solutions (\ref{51})
can be expressed in terms of the radial coordinate as follows
\begin{equation}
G_{j}=\widetilde{D}_{j}r^{i\widetilde{\alpha}}.\label{54}%
\end{equation}
An alternative way of getting the latter result is the decoupling of
$G_{j}$ from Eq. (\ref{38}):
\begin{equation}
\Lambda G_{j}^{\prime\prime}-\frac{b(r_{2}+r_{1}-2r)}{2}G_{j}^{\prime
}+\left\{  (-1)^{j}ik\left[  \frac{b(r_{2}r_{1}-r^{2})}{\Lambda}-\frac
{b(r_{2}+r_{1}-2r)}{2\Lambda}\right]  +\frac{k^{2}r^{2}}{\Lambda}%
-\frac{\lambda^{2}}{A^{2}}\right\}  G_{j}=0.\label{55}%
\end{equation}
As $r\rightarrow\infty$, Eq. (\ref{55}) takes the following form
\begin{equation}
r^{2}G_{j}^{\prime\prime}+rG_{j}^{\prime}+(\frac{k^{2}A^{2}-\lambda^{2}%
b}{A^{2}b^{2}}=\widetilde{\alpha}^{2})G_{j}=0.\label{56}%
\end{equation}
whose solutions are found to be
\begin{equation}
G_{j}=\widetilde{D}_{j}r^{i\widetilde{\alpha}}+\widehat{C}_{j}%
r^{-i\widetilde{\alpha}}.\label{57}%
\end{equation}
Since $e^{i\eta r_{\ast}}=r^{i\widetilde{\alpha}}$ and the wave should be
purely outgoing at the spatial infinity, we set $\widehat{C}_{j}=0$ and thus find
$G_{j}=\widetilde{D}_{j}r^{i\widetilde{\alpha}}$ which is nothing but the same
result that was obtained in Eq. (\ref{54}).

\section{Exact QNM frequencies}

\label{sec5}

To find the analytical solutions of $G_{j}$, we first introduce
\begin{equation}
G_{j}=H_{j}e^{(-1)^{j+1}ikr_{\ast}},\label{59}%
\end{equation}
and change the variable to
\begin{equation}
z=-x=-\frac{r-r_{2}}{r_{2}-r_{1}}. \label{60}%
\end{equation}
After inserting Eq. (\ref{60}) in Eq. (\ref{55}), we obtain
\begin{equation}
z(1-z)H_{j}^{\prime\prime}+\left[  \boldsymbol{c}-(1+\boldsymbol{a}%
+\boldsymbol{b})z\right]  H_{j}^{\prime}-\boldsymbol{a}\boldsymbol{b}%
H_{j}=0,\label{61}%
\end{equation}
which is the Euler's hypergeometric differential equation \cite{abram} with
\begin{align}
\boldsymbol{a}  & =i\left[  (-1)^{j+1}\frac{k}{b}+\widetilde{\alpha}\right]
,\nonumber\\
\boldsymbol{b}  & =i\left[  (-1)^{j+1}\frac{k}{b}-\widetilde{\alpha}\right]
,\nonumber\\
\boldsymbol{c}  & =\frac{1}{2}+2ik(-1)^{j+1}\frac{r_{2}}{b(r_{2}-r_{1}%
)}.\label{63}%
\end{align}
The general solution of Eq. (\ref{61}) is given by \cite{abram}
\begin{equation}
H_{j}=C_{2}F(\boldsymbol{a},\boldsymbol{b};\boldsymbol{c};z)+C_{1}%
z^{1-\boldsymbol{c}}F(\widetilde{\boldsymbol{a}},\widetilde{\boldsymbol{b}%
};\widetilde{\boldsymbol{c}};z),\label{64}%
\end{equation}
where $C_{1}$ and $C_{2}$ are the integration constants and $F(\boldsymbol{a},\boldsymbol{b}%
;\boldsymbol{c};z)$ and $F(\widetilde{\boldsymbol{a}}%
,\widetilde{\boldsymbol{b}};\widetilde{\boldsymbol{c}};z)$ are the
hypergeometric (or Gaussian) functions with
\begin{align}
\widetilde{\boldsymbol{a}}  & =\boldsymbol{a}-\boldsymbol{c}+1=\frac{1}%
{2}-i\left[  k(-1)^{j+1}\frac{r_{2}+r_{1}}{b(r_{2}-r_{1})}+\widetilde{\alpha
}\right]  ,\nonumber\\
\widetilde{\boldsymbol{b}}  & =\boldsymbol{b}-\boldsymbol{c}+1=\frac{1}%
{2}-i\left[  k(-1)^{j+1}\frac{r_{2}+r_{1}}{b(r_{2}-r_{1})}-\widetilde{\alpha
}\right]  ,\nonumber\\
\widetilde{\boldsymbol{c}}  & =2-\boldsymbol{c}=\frac{3}{2}-2ik(-1)^{j+1}%
\frac{r_{2}}{b(r_{2}-r_{1})}.\label{65}%
\end{align}
In the near-horizon region: $r\rightarrow r_{2},$ $r_{\ast}\rightarrow-\infty
$\ \ $\Rightarrow$\ \ $z\approx e^{r_{\ast}}\rightarrow0$. By recalling that
$F(\boldsymbol{a},\boldsymbol{b};\boldsymbol{c};0)=1$ \cite{abram}, the
physical solutions (\ref{59}), which fulfill the QNM condition that only
ingoing waves can propagate near the horizon, are found to be as follows
\begin{align}
G_{2}  & =H_{2}e^{-ikr_{\ast}}=C_{2}e^{-ikr_{\ast}}F(\boldsymbol{a}%
,\boldsymbol{b};\boldsymbol{c};z),\nonumber\\
G_{1}  & =H_{1}e^{ikr_{\ast}}=C_{1}z^{1-\boldsymbol{c}}e^{+ikr_{\ast}%
}F(\widetilde{\boldsymbol{a}},\widetilde{\boldsymbol{b}}%
;\widetilde{\boldsymbol{c}};z).\label{66}%
\end{align}
By using one of the special features of the hypergeometric functions
\cite{abram}:%
\begin{align}
F(\boldsymbol{a},\mathbf{b};\boldsymbol{c};y) &  =\frac{\Gamma(\boldsymbol{c}%
)\Gamma(\boldsymbol{b}-\boldsymbol{a})}{\Gamma(\boldsymbol{b})\Gamma
(\boldsymbol{c}-\boldsymbol{a})}(-y)^{-\boldsymbol{a}}F(\boldsymbol{a}%
,\boldsymbol{a}+1-\boldsymbol{c};\boldsymbol{a}+1-\boldsymbol{b}%
;1/y)\nonumber\\
&  +\frac{\Gamma(\boldsymbol{c})\Gamma(\boldsymbol{a}-\boldsymbol{b})}%
{\Gamma(\boldsymbol{a})\Gamma(\boldsymbol{c}-\boldsymbol{b})}%
(-y)^{-\boldsymbol{b}}F(\boldsymbol{b},\boldsymbol{b}+1-\boldsymbol{c}%
;\boldsymbol{b}+1-\boldsymbol{a};1/y),\label{67}%
\end{align}
one can extend the solutions (\ref{66})\ to spatial infinity ($r\rightarrow\infty,$
$r_{\ast}\rightarrow\infty$ $\Rightarrow\ \frac{1}{z}\rightarrow0$) as follows%
\begin{align}
G_{1}  & =C_{1}z^{1-\boldsymbol{c}}e^{ikr_{\ast}}\textcolor{black}{\bigg[}\frac{\Gamma
(\widetilde{\boldsymbol{c}})\Gamma(\widetilde{\boldsymbol{b}}-\widetilde{\boldsymbol{a}}%
)}{\Gamma(\widetilde{\boldsymbol{b}})\Gamma(\boldsymbol{c}-\widetilde{\boldsymbol{a}})}(x)^{-\widetilde{\boldsymbol{a}}%
}F(\widetilde{\boldsymbol{a}},\widetilde{\boldsymbol{a}}+1-\widetilde{\boldsymbol{c}};\widetilde{\boldsymbol{a}}+1-\widetilde{\boldsymbol{b}}%
;1/z)\nonumber\\
& +\frac{\Gamma(\widetilde{\boldsymbol{c}})\Gamma(\widetilde{\boldsymbol{a}}-\widetilde{\boldsymbol{b}})}%
{\Gamma(\widetilde{\boldsymbol{a}})\Gamma(\widetilde{\boldsymbol{c}}-\widetilde{\boldsymbol{b}})}(x)^{-\widetilde{\boldsymbol{b}}%
}F(\widetilde{\boldsymbol{b}},\widetilde{\boldsymbol{b}}+1-\widetilde{\boldsymbol{c}};\widetilde{\boldsymbol{b}}+1-\widetilde{\boldsymbol{a}}%
;1/z)\textcolor{black}{\bigg],}\nonumber\\
G_{2}  & =C_{2}e^{-ikr_{\ast}}\textcolor{black}{\bigg[}\frac{\Gamma(\boldsymbol{c})\Gamma(\boldsymbol{b}-\boldsymbol{a})}{\Gamma
(\boldsymbol{b})\Gamma(\boldsymbol{c}-\boldsymbol{a})}(x)^{-\boldsymbol{a}}F(\boldsymbol{a},\boldsymbol{a}+1-\boldsymbol{c};\boldsymbol{a}+1-\boldsymbol{b};1/z)\nonumber\\
& +\frac{\Gamma(\boldsymbol{c})\Gamma(\boldsymbol{a}-\boldsymbol{b})}{\Gamma\boldsymbol{a})\Gamma(\boldsymbol{c}-\boldsymbol{b})}(x)^{-\boldsymbol{b}}%
F(\boldsymbol{b},\boldsymbol{b}+1-\boldsymbol{c};\boldsymbol{b}+1-\boldsymbol{a};1/z)\textcolor{black}{\bigg]}.\label{68}%
\end{align}
Since $x=\frac{r-r_{2}}{r_{2}-r_{1}},$ and thus $x\simeq r\simeq
e^{br_{\ast}}$ at the infinity, we have%
\begin{align}
G_{1}  & \approx C_{1}\frac{\Gamma(\widetilde{\boldsymbol{c}})\Gamma(\widetilde{\boldsymbol{b}}%
-\widetilde{\boldsymbol{a}})}{\Gamma(\widetilde{\boldsymbol{b}})\Gamma(\boldsymbol{c}-\widetilde{\boldsymbol{a}})}%
r^{i\widetilde{\alpha}}+C_{1}\frac{\Gamma(\widetilde{\boldsymbol{c}})\Gamma(\widetilde{\boldsymbol{a}}%
-\widetilde{\boldsymbol{b}})}{\Gamma(\widetilde{\boldsymbol{a}})\Gamma(\widetilde{\boldsymbol{c}}-\widetilde{\boldsymbol{b}}%
)}r^{-i\widetilde{\alpha}}. \label{69x}\\
G_{2}  & \approx C_{2}\frac{\Gamma(\boldsymbol{c})\Gamma(\boldsymbol{b}%
-\boldsymbol{a})}{\Gamma(\boldsymbol{b})\Gamma(\boldsymbol{c}-\boldsymbol{a}%
)}r^{i\widetilde{\alpha}}+C_{2}\frac{\Gamma(\boldsymbol{c})\Gamma
(\boldsymbol{a}-\boldsymbol{b})}{\Gamma(\boldsymbol{a})\Gamma(\boldsymbol{c}%
-\boldsymbol{b})}r^{-i\widetilde{\alpha}}.\label{69}%
\end{align}
The correspondence between the asymptotic solutions (\ref{50}) and the above
solutions (\ref{69}) yields%
\begin{align}
\widetilde{D}_{1}  & =C_{1}\frac{\Gamma(\widetilde{\boldsymbol{c}})\Gamma(\widetilde{\boldsymbol{b}}%
-\widetilde{\boldsymbol{a}})}{\Gamma(\widetilde{\boldsymbol{b}})\Gamma(\boldsymbol{c}-\widetilde{\boldsymbol{a}})},\nonumber\\
\widetilde{D}_{2}  & =C_{2}\frac{\Gamma(\boldsymbol{c})\Gamma(\boldsymbol{b}-\boldsymbol{a})}{\Gamma(\boldsymbol{b})\Gamma
(\boldsymbol{c}-\boldsymbol{a})},\nonumber\\
\widetilde{D}_{3}  & =C_{1}\frac{\Gamma(\widetilde{\boldsymbol{c}})\Gamma(\widetilde{\boldsymbol{a}}%
-\widetilde{\boldsymbol{b}})}{\Gamma(\widetilde{\boldsymbol{a}})\Gamma(\widetilde{\boldsymbol{c}}-\widetilde{\boldsymbol{b}}%
)},\nonumber\\
\widetilde{D}_{4}  & =C_{2}\frac{\Gamma(\boldsymbol{c})\Gamma(\boldsymbol{a}%
-\boldsymbol{b})}{\Gamma(\boldsymbol{a})\Gamma(\boldsymbol{c}-\boldsymbol{b}%
)}.\label{70}%
\end{align}
Therefore, to have only pure outgoing waves at the spatial infinity
(i.e., $\widetilde{D}_{3},\widetilde{D}_{4}=0$), we appeal to the pole structure of
the Gamma functions. The Gamma functions
$\Gamma(\widetilde{x})$ have poles at $\widetilde{x}=-n$ for $n=0,1,2$....
\cite{abram}. Hence, we obtain the QNMs by using the following relations:%
\begin{align}
\boldsymbol{a}  & =i\left[  -\frac{k}{b}+\widetilde{\alpha}\right]  =-n,\nonumber\\
\boldsymbol{c}-\boldsymbol{b}  & =\frac{1}{2}+i\left[  -k\frac{r_{2}+r_{1}}{b(r_{2}-r_{1})}%
-\widetilde{\alpha}\right]  =-n,\nonumber\\
\widetilde{\boldsymbol{a}}  & =\frac{1}{2}-i\left[  k\frac{r_{2}+r_{1}}{b(r_{2}-r_{1}%
)}+\widetilde{\alpha}\right]  =-n,\nonumber\\
\widetilde{\boldsymbol{c}}-\widetilde{\boldsymbol{b}}  & =1+i\left[  k\frac{r_{2}+r_{1}}{b(r_{2}-r_{1}%
)}-\widetilde{\alpha}\right]  =-n.\label{71}%
\end{align}
Recall that the time dependence of the QNMs is governed by $e^{ikt}$
(see Eq.(\ref{22})). Therefore, having QNMs is conditional on $Im(k)>0$, which
guaranties the stability \cite{isp40}. The conditions under which frequencies obtained from $G_{1}$ and $G_{2}$ spinor solutions produce stable modes (i.e., QNMs) and which ones generate the unstable modes are summarized in Tables \ref{tab1} and \ref{tab2}:
\begin{table}[H]
\begin{tabular}
[c]{|l|l|l|}\hline
Frequencies & Stable Modes (QNMs) & Unstable Modes\\\hline%
\begin{tabular}
[c]{l}%
$k=-\frac{I}{2A(\beta^{2}-1)}\left[  A\beta b(2n+1)+\right.  $\\
$\left.  2\sqrt{b}\sqrt{A^{2}b(n+\frac{1}{2})^{2}+\lambda^{2}(\beta^{2}%
-1))}\right]  $%
\end{tabular}
& if $A<0$ and $A<\frac{\lambda}{\sqrt{b}(n+\frac{1}{2})}$ &
\begin{tabular}
[c]{l}%
if $A>0$ or\\
if $A<0$ and $A>\frac{\lambda}{\sqrt{b}(n+\frac{1}{2})}$%
\end{tabular}
\\\hline%
\begin{tabular}
[c]{l}%
$k=\frac{I}{2A(\beta^{2}-1)}\left[  A\beta b(n+1)+\right.  $\\
$\left.  \sqrt{b}\sqrt{A^{2}b(n+1)^{2}+\lambda^{2}(\beta^{2}-1))}\right]  $%
\end{tabular}
&
\begin{tabular}
[c]{l}%
if $A>0$\\
if $A<0$ and $A<\frac{\lambda}{\sqrt{b}(n+1)}$%
\end{tabular}
& if $A<0$ and $A>\frac{\lambda}{\sqrt{b}(n+1)}$\\\hline
\end{tabular}
\caption{QNM\ frequencies for $G_{1}$. The results are obtained from Eqs. (\ref{69x}) and (\ref{71}).} \label{tab1}
\end{table}

\begin{table}[H]
\begin{tabular}
[c]{|l|l|l|}\hline
Frequencies & Stable Modes (QNMs) & Unstable Modes\\\hline%
$k=\frac{I}{2A^{2}n}(\lambda^{2}-n^{2}A^{2}b)$ &
\begin{tabular}
[c]{l}%
if $A>0$ and $A<\frac{-\lambda}{\sqrt{b}n}$\\
if $A<0$ and $A>\frac{\lambda}{\sqrt{b}n}$%
\end{tabular}
&
\begin{tabular}
[c]{l}%
if $A>0$ and $A>\frac{-\lambda}{\sqrt{b}n}$\\
if $A<0$ and $A<\frac{\lambda}{\sqrt{b}(n}$%
\end{tabular}
\\\hline%
\begin{tabular}
[c]{l}%
$k=-\frac{I}{2A(\beta^{2}-1)}\left[  A\beta b(2n+1)+\right.  $\\
$\left.  2\sqrt{b}\sqrt{A^{2}b(n+\frac{1}{2})^{2}+\lambda^{2}(\beta^{2}%
-1))}\right]  $%
\end{tabular}
& if $A<0$ and $A<\frac{\lambda}{\sqrt{b}(n+\frac{1}{2})}$ &
\begin{tabular}
[c]{l}%
if $A>0$ or\\
if $A<0$ and $A>\frac{\lambda}{\sqrt{b}(n+\frac{1}{2})}$%
\end{tabular}
\\\hline
\end{tabular}
\caption{QNM\ frequencies for $G_{2}$. The results are obtained from Eqs. (\ref{69}) and (\ref{71}).} \label{tab2}
\end{table}
In Tables \ref{tab1} and \ref{tab2}, the parameter of $\beta$ is given by
\begin{equation}
1<\beta=\frac{r_{2}+r_{1}}{r_{2}-r_{1}}<\infty.\label{72}%
\end{equation}
As can be seen from above, depending on the values of $A,\lambda,b,n$
some QNM frequencies can have positive/negative imaginary part, and therefore
the LDBHs can be stable/unstable under charged fermionic perturbations.
\section{\bigskip Conclusion} \label{sec6}

In this paper, we have analytically computed the QNMs of charged fermionic
perturbations for the $4$-dimensional LDBHs which are the solutions to the EMD
theory with the double Lioville-type potentials. The fermionic fields are the solutions of the Dirac equation. For the LDBH geometry, we have found that the massless but charged fermionic fields are governed by the Gaussian hypergeometric functions. We matched the exact solutions obtained with the solutions found for near-horizon and asymptotic regions. Thus, we have shown that there are two sets of frequencies for each component ($G_{1}$ and $G_{2}$) of the fermionic fields. As being summarized in Tables \ref{tab1} and \ref{tab2}, in each set the frequencies obtained can belong to either the stable modes (QNMs) or the unstable modes depending on the relations between the parameters of $A,\lambda,b,$ and $n$. Meanwhile, the reader may question what the advantage/disadvantage of perturbing the LDBH with fermionic fields instead of the bosonic fields is. If precise measurements of QNM frequencies become possible in the future and if the fermionic QNM data obtained for spin-up and spin-down fields (i.e., $G_{1}$ and $G_{2}$) coincide with (a kind of double check) the analytical expressions that we have derived, those results will likely reveal the presence of LDBHs more accurately than the bosonic QNM ones.

As is known, rotating BH solutions are more considerable for testing relevant theories with the astrophysical observations. For this reason, in the near future, we plan to extend our study to the rotating LDBHs. To this end, we shall apply the Newman–Janis algorithm \cite{ispson} to the metric (\ref{11}) and perturb the obtained rotating LDBH to reveal the effect of rotation on their QNMs. 

\section*{Acknowledgements}

The authors are grateful to the Editor and anonymous Referees for their valuable comments and suggestions to improve the paper.

\end{document}